\documentclass[a4paper,11pt]{article}
\usepackage{pos}
\usepackage{amsmath}
\usepackage[export]{adjustbox}

\title{Shear viscosity from quenched to full lattice QCD}
\author*[a]{Pavan}
\author[a]{Olaf Kaczmarek}
\author[b]{Guy D. Moore}
\author[a]{Christian Schmidt}

\affiliation[a]{Fakult\"at f\"ur Physik, Universit\"at Bielefeld, D-33615 Bielefeld, Germany}
\affiliation[b]{Institut für Kernphysik, Technische Universität Darmstadt, D-64289 Darmstadt, Germany}

\abstract{
The shear viscosity of the quark-gluon plasma (QGP) plays a crucial role in interpreting current measurements from heavy-ion collisions and is a key input to hydro-dynamical models. The interest in shear viscosity also lies in the fact that QGP is the most ideal fluid ever observed and has the shear viscosity to entropy ratio ($\eta / s$) close to the theoretical bound $\eta / s \geq 1/ 4 \pi$ in the strong coupling region within AdS/CFT formalism. The lattice determination of $\eta / s$ has been explored for the pure gauge case, but its determination in full QCD remains unexplored, despite its significant importance. In this proceeding, we present updates on extending our quenched findings from~\cite{LA} to full QCD. Specifically, we focus on the renormalization of the energy-momentum tensor with the gradient flow method and provide a progress update on determining the relevant renormalization coefficients for shear viscosity. For this purpose, we have used an imaginary isospin chemical potential.
}

\FullConference{The 41st International Symposium on Lattice Field Theory (LATTICE2024)\\
 28 July - 3 August 2024\\
Liverpool, UK\\}

\begin{document}
\maketitle
\section{Introduction}
Shear viscosity quantifies the resistance of hydrodynamic systems to shear flow near equilibrium. It plays a critical role in hydrodynamical models and is associated with the off-diagonal component of the leading-order correction to the equilibrium energy-momentum tensor:
\begin{equation}
        T_{ij} - T^{eq}_{ij} = -\eta \, \Big( \nabla_{i} u_{j} + \nabla_{j} u_{i} - \frac{2}{3} \, \delta_{ij} \, \nabla_{k} u_{k} \Big) - \zeta \, \delta_{ij} \, \nabla_{k} u_{k}.
\end{equation}
Here $\eta$ and $\zeta$ are shear and bulk viscosities respectively.

The experimental studies \cite{KHA,SSA,KA,GA,JA} along with lattice pure-glue studies~\cite{LA,HM,FW} show a small shear viscosity to entropy ratio ($\eta /s $) which is close to predicted lower bound $1/4\pi$ from AdS/CFT~\cite{GD}.

Perturbation theory has limitations in determining shear viscosity because potential energies are of the same order as kinetic energies and temperature, as discussed in~\cite{GM}. For instance, Figure~1 in~\cite{GJ} shows that even at $T \sim 100 $ GeV the shear viscosity to entropy density ratio ($\eta /s $) at leading order is twice that at next-to-leading order. This discrepancy highlights the necessity of using lattice methods to determine shear viscosity.

The shear viscosity is given by following Kubo relation in real-time:
\begin{equation}
        \eta = i\partial_\omega \int d^3 x \int_0^\infty dt \,e^{i \omega t} \langle  [T^{xy}(x,t),T^{xy}(0,0)] \rangle\big\vert_{\omega = 0}.
\end{equation}
The above real-time commutation relation is not accessible on the lattice; instead, we can only evaluate the Euclidean time correlation function. This Euclidean correlation function is related to shear viscosity through analytical continuation in terms of the corresponding spectral function, $\rho_{\mathrm{shear}}(\omega,T)$, via
\begin{equation}
        \eta(T) =\lim_{\omega\rightarrow  0} \frac{\rho_{\mathrm{shear}}(\omega,T)}{\omega}.
\end{equation}
The spectral function can not be computed directly on the lattice but is related to the correlation function by following relation:
\begin{equation}
        G(\tau)=\int_0^{\infty}\frac{\mathrm{d}\omega}{\pi}\frac{\cosh[\omega(1/2T-\tau)]}{\sinh(\omega/2T)}\rho(\omega,T).
\end{equation}
Only a finite number of data points are available for lattice correlators, $G(\tau)$, which permits many possible spectral functions to fit within the error bars. This is referred to as an ill-conditioned inversion problem, where the goal is to determine the spectral function $\rho(\omega)$ from limited and incomplete information. To address this, we model the spectral function using input from physics and extract the shear viscosity by fitting the lattice correlators.

Another challenge is the lattice lacks continuous space-time translational symmetry, making the choice and renormalization of the energy-momentum tensor neither obvious nor unique. To address this, we used gradient flow to construct the energy-momentum tensor operators on the lattice and to reduce the noise arising from their correlations. These issues have already been addressed in our previous work on the pure-glue case, published in ~\cite{LA}.

In this proceeding, we provide updates on extending our viscosity determination method to full QCD, which is considerably more complex due to various technical challenges. One first major challenge involves renormalizing the energy-momentum tensor and determining the associated renormalization coefficients. In the case of pure glue, we utilized the enthalpy density to fix these coefficients. However, in full QCD, an additional renormalization coefficient arises from the fermionic contribution to the shear viscosity, requiring two independent equations to determine these coefficients. Identifying these independent equations is a challenging task, which we will discuss in detail in a later section of this proceeding.

\section{Theoretical Background}
The inclusion of fermions in the lattice determination of shear viscosity has remained unexplored due to technical complexities. In full QCD, the energy-momentum tensor contains the following five terms under $SO(4)$ symmetry \cite{HH},
  \begin{align}
   T^1_{\mu\nu}(x, \tau_F) &\equiv Z_1(\tau_F) \Big [ F_{\mu\rho}^a(x, \tau_F)F_{\nu\rho}^a(x, \tau_F) - \frac{1}{4} \delta_{\mu \nu} F_{\rho \sigma}^a(x, \tau_F)F_{\rho \sigma}^a(x, \tau_F) \Big ]
\\
   T^2_{\mu\nu}(x, \tau_F) &\equiv Z_2(\tau_F) \,
   \delta_{\mu\nu}F_{\rho\sigma}^a(x, \tau_F)F_{\rho\sigma}^a(x, \tau_F)
\\
   T^3_{\mu\nu}(x, \tau_F) &\equiv Z_3(\tau_F) \, \bar{\psi}(x)
   \Big [ \gamma_\mu\overleftrightarrow{D}_\nu
   +\gamma_\nu\overleftrightarrow{D}_\mu - \frac{1}{2} \delta_{\mu \nu} \gamma_\alpha\overleftrightarrow{D}_\alpha \Big ]
   \psi(x)
\\
   T^4_{\mu\nu}(x, \tau_F) &\equiv Z_4(\tau_F) \,
   \delta_{\mu\nu}
   \bar{\psi}(x) \gamma_\alpha
   \overleftrightarrow{D}_\alpha
   \psi(x)
\\
   T^5_{\mu\nu}(x, \tau_F) &\equiv Z_5 (\tau_F) \,
   \delta_{\mu\nu}
   m_0 \bar{\psi}(x)
   \psi(x).
\end{align}
We are only applying the gradient flow to the gauge fields, not explicitly to the fermionic fields (mixed gradient flow). Among the above terms, only the terms $T^1_{\mu\nu}(x, \tau_F)$ and $T^3_{\mu\nu}(x, \tau_F)$ are associated to shear viscosity, necessitating the determination of the corresponding renormalization coefficients $Z_1$ and $Z_3$. Determining these renormalization coefficients is challenging because the method we employed in \cite{LA}, which was inspired by the work in \cite{GP}, requires two independent pieces of information to find the two renormalization coefficients $Z_1$ and $Z_3$.

To address this, we follow ideas first developed by Giusti and Pepe \cite{GP2,GP3,GP4}.
We require two distinct ensembles to extract the two renormalization coefficients using the relation $\epsilon + p = \langle T_{xx} - T_{00} \rangle$. The absence of full $SO(4)$ symmetry on the lattice causes the renormalization coefficients for $T_{xx} -T_{00}$ to differ from those for $T_{xy}$.
However, the use of gradient flow will suppress this effect at the level of $\mathcal{O}(a^2/\tau_F)$, and provided that one takes first a continuum and then a small $\tau_F$ limit, this complication goes away.

We obtain $\epsilon + p$ by measuring the interaction measure and pressure,  as detailed in the latter part of this proceeding. The quantity $\langle T_{xx} - T_{00} \rangle $ is determined by evaluating the following operator:
\begin{align}
(T^G_{xx} - T^G_{00}) (x, \tau_F)&\equiv \frac{Z_1(\tau_F)}{g_0^2} \big (F_{x\alpha}(x, \tau_F) F_{x\alpha}(x, \tau_F)  - F_{0\alpha}(x, \tau_F) F_{0\alpha}(x) \big)
\\
 (T^F_{xx} - T^F_{00})(x, \tau_F) &\equiv \frac{Z_3(\tau_F)}{2} \bar{\psi}(x) \big(\gamma_x \overleftrightarrow{D}_x - \gamma_0 \overleftrightarrow{D}_0 \big) \psi(x) 
\end{align}
The above operators are associated with physical observables, so they must remain finite in the zero-flow-time limit. However, individually, the renormalization coefficients or the bare operators may diverge, as was the case in pure-glue \cite{LA}. In the next section, we will outline our approach to determining these renormalization coefficients.

\section{Free theory motivation to get independent equation}

The idea behind getting two independent constraints is, reducing the fermionic pressure contribution significantly without much affecting the gluonic contribution. This can be achieved by using imaginary isospin chemical potential. In the presence of an imaginary chemical potential, the phase structure is governed by center symmetry, even though the presence of fermions breaks this symmetry explicitly ~\cite{RW}. This leads to a phase transition at $\mu^u_I/T = \mu^d_I/T = \mu^s_I/T = \pi/3$, known as the Roberge-Weiss phase transition. At this point, the fermionic contribution to the pressure is reduced by a factor of less than 2. Our proposed approach involves setting $\mu^u_I/T = - \mu^d_I/T = \theta $ and $\mu^s_I/T = 0$.  In the right plot of Figure~1, we have shown the pressure corresponding to different Polyakov loop vacua, which demonstrates that a phase transition occurs at $\theta = 2 \pi/3$. At this point, the fermionic pressure is suppressed by a factor of 81 in the free massless theory.

\begin{figure}[h!]
\centering
\includegraphics[width=0.41\textwidth]{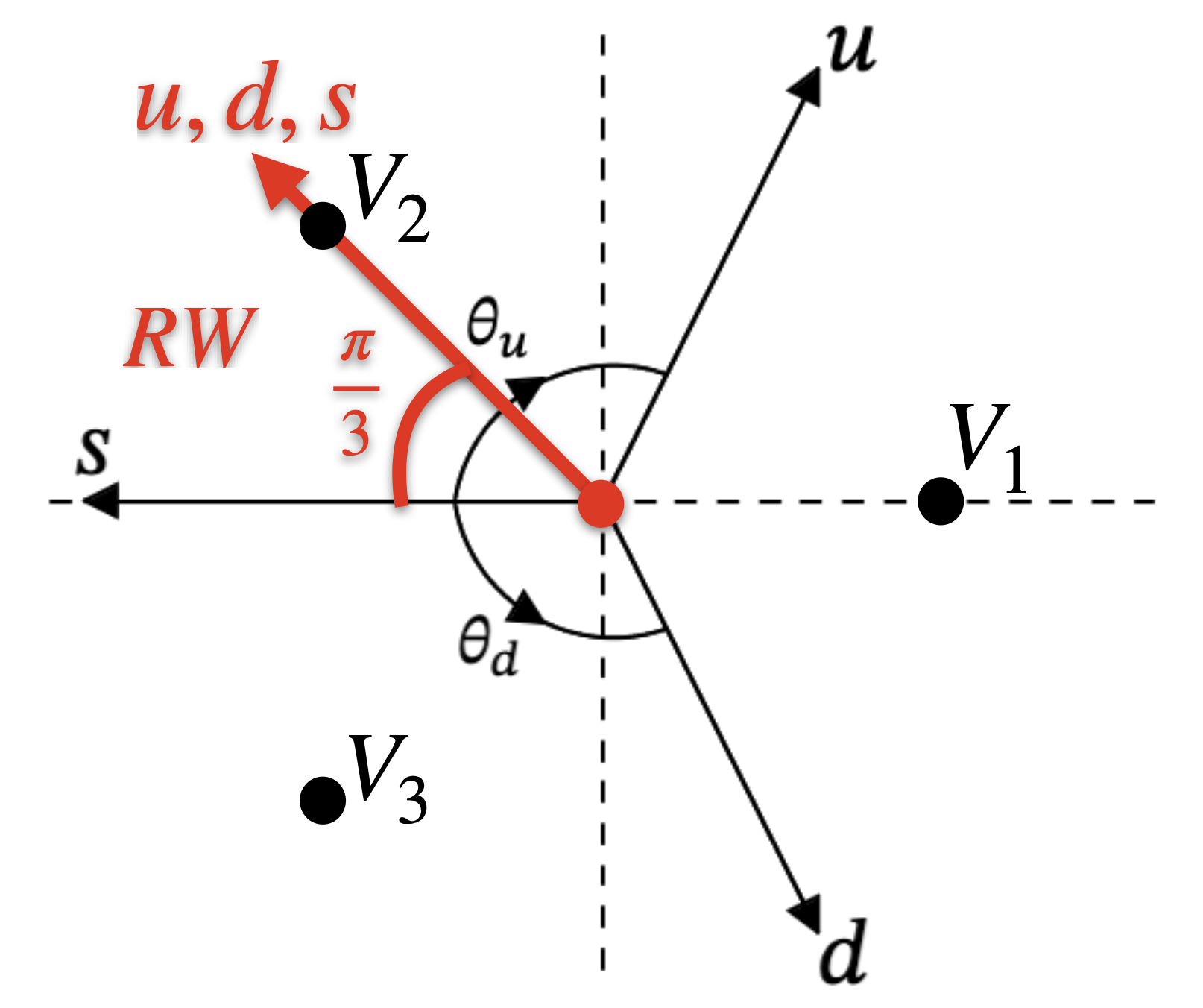}
\includegraphics[width=0.49\textwidth]{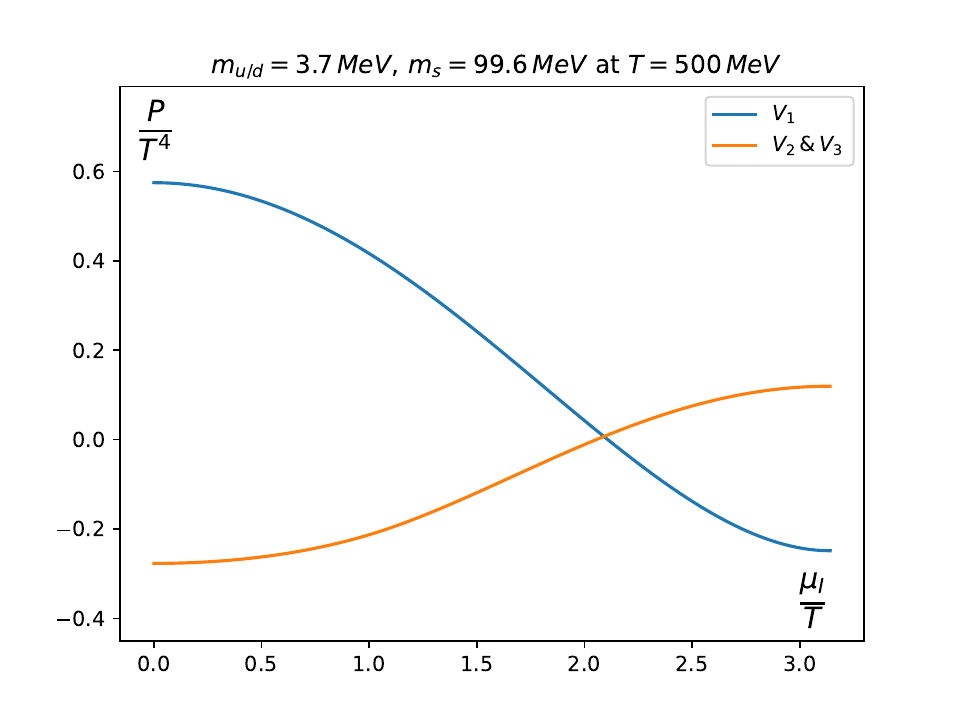}
\caption{(Left) Possible Polyakov loop vacua $V_{1,2,3}$, and the quark twist angles for the Roberge-Weiss case (red) and our case (black).
(Right) Free fermionic pressure for the Polyakov-loop vacua $V_1$, $V_2$ and $V_3$. }
\label{lattice-corr-sp}
\end{figure}

The above observation is based on free theory calculations, and it is important to understand how it behaves when interactions are included. We expect this behavior to persist, at least at high temperatures, as interactions play a less important role with increasing temperature.

So we will do our analysis for $\epsilon + p = \langle T_{xx} - T_{00} \rangle$ in the following ensembles,\\
Ensemble 1: ~~~~~~~$\mu_I^u /T = \mu^d_I /T = \mu^s_I /T = 0 ~~(\equiv \mu_i)$ and  \\
Ensemble 2: ~~~~~~~$\mu^u_I/T  = -\mu^d_I/T  = 2 \pi /3$ and $\mu^s_I/T = 0 ~~(\equiv \mu_f)$ \\
\\
The measurement of $\epsilon + p$ at $\mu_f$ requires determining both the interaction measure and the pressure at $\theta = 2 \pi / 3$. While the interaction measure is directly accessible on the lattice, measuring the pressure necessitates evaluating the number density at various values of $\theta$, integrating the number density yields the pressure. Further, we need to measure the expectation values of the operators in Equation (10) and Equation (11) on the above two ensembles. We expect that the gluonic contribution in Equation (10) is not highly sensitive to the imaginary chemical potential, and the overall suppression of $\epsilon + p$ is dominated by the fermionic contribution in Equation (11).

\section{Result and Analysis}
In this section, we present our results and the details involved. The analysis is performed for $\beta = 7.373$ on a $40^3 \times 8$ Euclidean lattice at $T = 409.7$ MeV and $m_l = m_s/20$. The configurations are generated using the HISQ action for $1 + 1 + 1$ flavors, where the breaking between the $u$ and $d$ quarks arises due to different imaginary isospin chemical potentials. We determine the entropy density at $\theta = 0$ and $\theta = 2 \pi/3$ by measuring the interaction measure and pressure, $\epsilon + p = I + 4p$. The interaction measure is given as:
\begin{equation}
\frac{I}{T^4}  
= R_\beta \Big [
\Big ( \langle S_G \rangle_0 - \langle S_G \rangle_\tau \Big ) - R_{m} \sum_{q = u, d, s} 
 m_{q}\Big( \langle\bar{\psi}\psi \rangle_{q,0}
- \langle\bar{\psi}\psi \rangle_{q,\tau} \Big)
 \Big] N_\tau^4
 \end{equation}
 The nonperturbative beta function and mass renormalization function are defined as \cite{AB}
\begin{equation}
 R_{\beta}(\beta) = \frac{r_1}{a} \left( \frac{{\rm d} (r_1/a)}{{\rm d} \beta} \right)^{-1}  ~~~~~~~ R_m(\beta) = \frac{1}{m_s(\beta)}
\frac{{\rm d} m_s(\beta)}{{\rm d}\beta} 
 \end{equation}
\begin{equation}
\left . r^2 \frac{dV}{dr} \right|_{r_i}=C_i,~\;\; i=0,\,1;  ~~~~~~~~~~~~~~~~~ V(r) = C + \frac{B}{r} + \sigma r
\end{equation}

The zero-temperature values for renormalization are taken from reference~\cite{AB} which has same $\beta$ and temperature, although the spatial volume is not exactly the same.

\begin{figure}[h]
\centering
\includegraphics[width=0.48\textwidth]{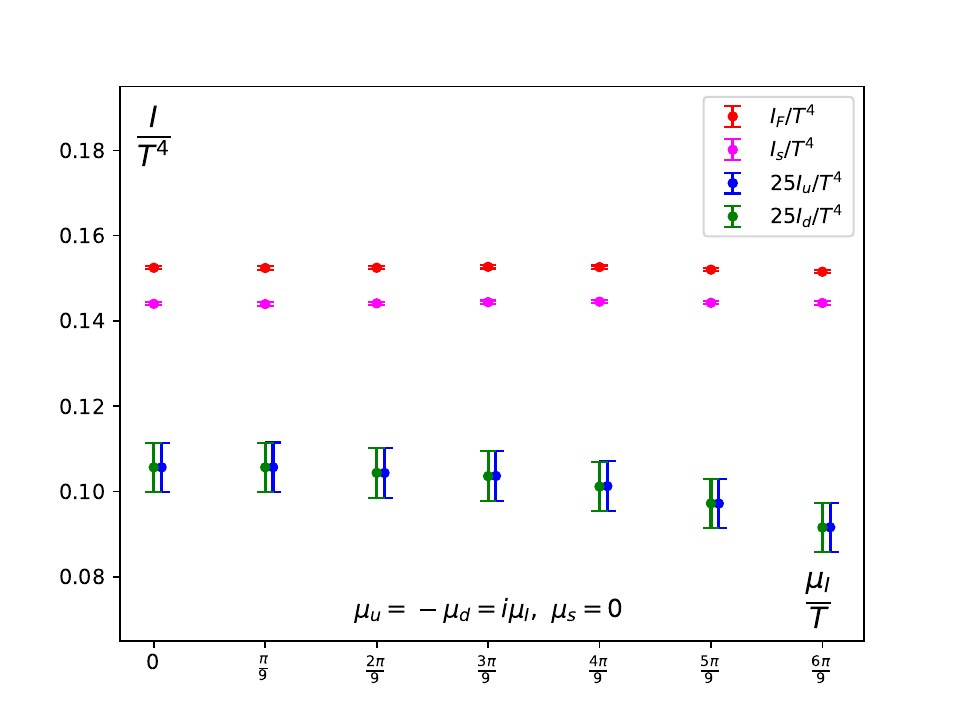}
\includegraphics[width=0.48\textwidth]{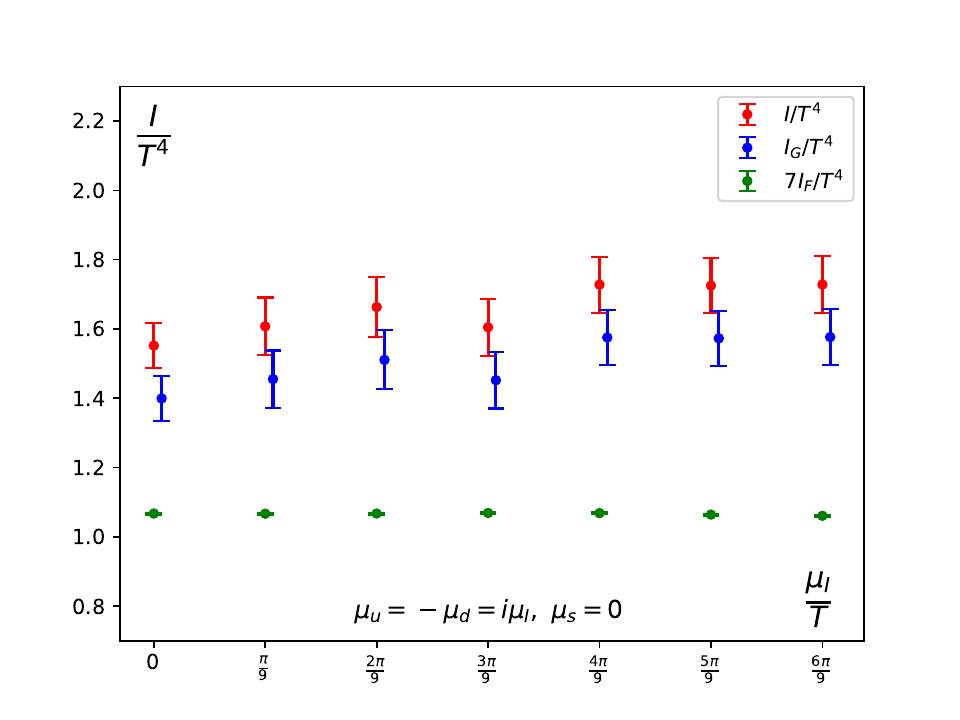}
\caption{(Left) The fermionic contribution for trace anomaly.
(Right) Gluonic and fermionic contribution to trace anomaly. }
\label{lattice-corr-sp2}
\end{figure}

We observed that the overall interaction measure does not significantly depend on the imaginary isospin chemical potential. This is because the dominant contribution to the interaction measure originates from the gluonic and strange quark sectors, which are not explicitly affected by the imaginary isospin chemical potential.

The next step involves calculating the pressure at $\theta = 0$ and $\theta = 2 \pi/3$. Since the pressure is not directly accessible on the lattice, we determine it by measuring the number density at intermediate values of the $\theta$ and then integrating it over the imaginary isospin chemical potential. The number density for an individual fermionic component is expressed as:
\begin{equation}
n = \frac{1}{4} \frac{T}{V} \, \bigg \langle \text{Tr} \Big [ M(\mu)^{-1} \frac{\partial M(\mu) }{\partial \mu}  \Big ] \bigg \rangle.
\end{equation}
We used the following fit function for the number density, motivated by the fact that the partition function is an even function of the chemical potential.
\begin{equation}
    \frac{n}{T^3} = a\left(\frac{\mu}{T} \right) + b \left(\frac{\mu}{T} \right)^3 + c\left(\frac{\mu}{T} \right)^5 + d\left(\frac{\mu}{T} \right)^7.
\end{equation}
The pressure will be obtained by integrating the number density as follows:
\begin{equation}
n = \frac{N}{V} = \frac{\partial p}{\partial \mu} \Big |_{T} ~~~~\Rightarrow ~~~~ p (\mu_f) = p(\mu_I = 0) + \int^{\mu_f}_{0} \frac{\partial p}{\partial \mu_I} \, d\mu_I.
\end{equation}
Here $p(\mu_I=0)$ the zero-$\mu$ pressure is extracted from data published in \cite{AB} and is $p_0/T^4 = 3.836909 \pm 0.245714$.  

\begin{figure}[h]
\centering
\includegraphics[width=0.48\textwidth]{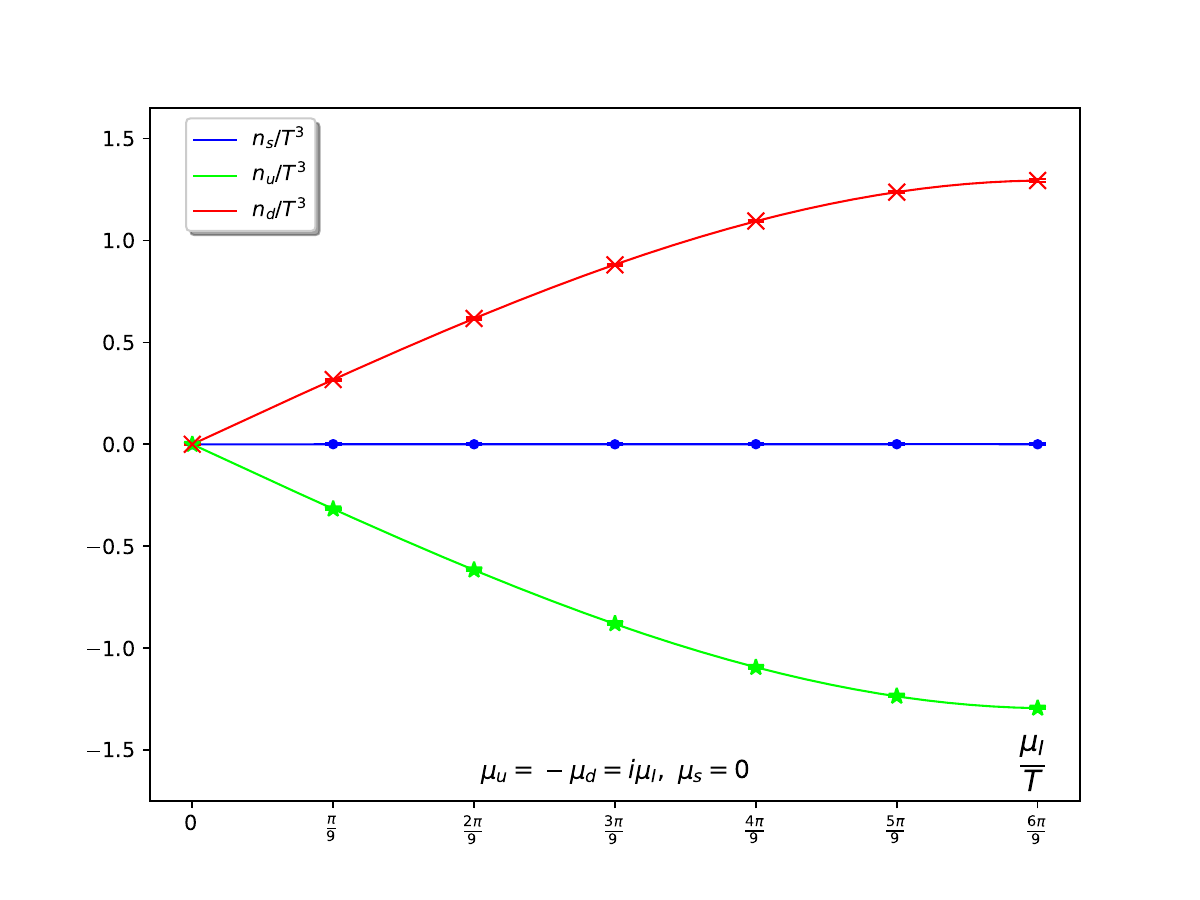}
\includegraphics[width=0.48\textwidth]{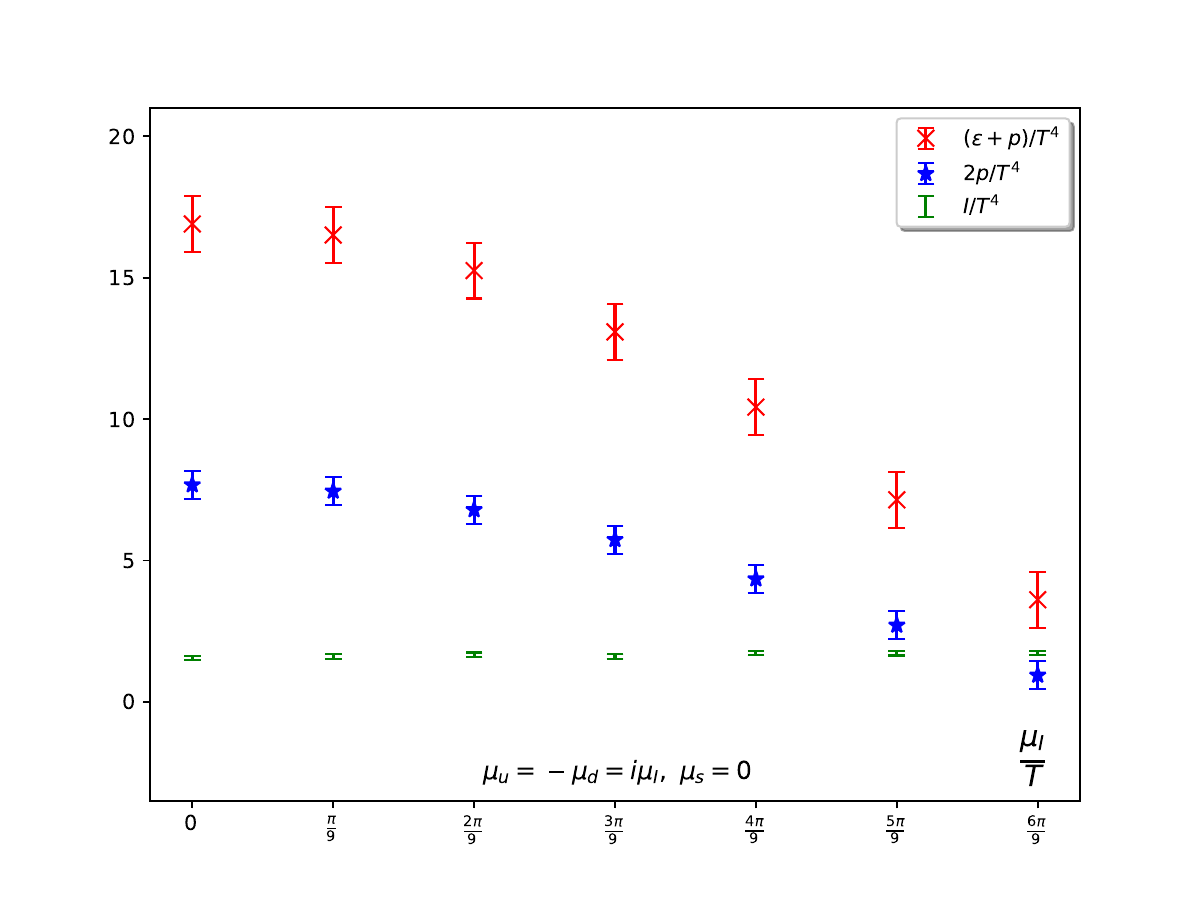}
\caption{(Left) The number density plot for $u$, $d$ and $s$ flavors.
(Right) The enthalpy density with imaginary isospin chemical potential. }
\label{lattice-corr-sp3}
\end{figure}

The major contribution to the error in enthalpy density arises from the zero-$\mu$ pressure, which is challenging to control as we rely on values reported in literature studies. Improving this is computationally expensive; for instance, it would require integrating over the mass from the quenched theory at the same $\beta$ value. We observed that the pressure decreases significantly with increasing $\theta$, consistent with our free theory predictions. However, the reduction factor is comparatively smaller than in the case of free theory, indicates interactions are playing significant role in the dynamics.

It is evident from the right plot of Figure~3 that $\epsilon+p$ differs significantly between $\theta = 0$ and $\theta = 2 \pi /3$. This provides a clear indication that we will obtain an independent equation, as the presence of the imaginary isospin chemical potential does not alter the gluonic contribution at leading order. The observed difference mostly arises from the fermionic contribution.

\section{Conclusion and Outlook}
We presented an update on determining the renormalization coefficients relevant for calculating shear viscosity. Using the gradient flow method, we constructed the renormalized energy-momentum tensor on the lattice and developed a framework for determining the corresponding renormalization coefficients. Thus far, we have calculated the enthalpy density and are currently working on evaluating the expectation values of the relevant energy-momentum tensor operators. The significant dependence of the enthalpy density on the imaginary isospin chemical potential indicates that our method will likely yield the renormalization coefficients. The major part of error in the enthalpy density arises from the pressure at zero-$\mu$. We are working to mitigate or bypass it by effectively utilizing the available information. Additionally, we are extending this study to two additional $\beta$ values to achieve the continuum limit. Once these renormalization coefficients are obtained, we will follow an approach similar to that explored in~\cite{LA}, while addressing the additional challenges arising in the process.

\acknowledgments
We thank Dibyendu Bala, Michele Pepe and Peter Petreczky for stimulating and helpful discussions. The authors acknowledge the support by the Deutsche Forschungsgemeinschaft (DFG, German Research Foundation) through the CRC-TR 211 'Strong-interaction matter under extreme conditions'– project number 315477589 – TRR 211, and PUNCH4NFDI with grand 460248186. The computations have been performed using the LUMI-G supercomputer and the GPU-cluster facilities at Bielefeld University, we thank the Bielefeld HPC.NRW team for their support. We acknowledge the EuroHPC Joint Undertaking for awarding this project access to the EuroHPC supercomputer LUMI-G, hosted by CSC (Finland) and the LUMI consortium through a EuroHPC Extreme Scale Access call.\\


\begin{thebibliography}{99}
\bibitem{LA}
L. Altenkort \textit{et al.}, \emph{Viscosity of pure-glue QCD from the lattice}, \href{https://doi.org/10.1103/PhysRevD.108.014503}{Phys. Rev. D 108 (2023) 014503}, [\href{https://arxiv.org/abs/2211.08230}{\tt{2211.08230}}].

\bibitem{KHA}
K. H. Ackermann et al. (STAR Collaboration), \emph{Elliptic Flow in $Au + Au$ Collisions at $\sqrt{s_{\text{NN}}}$ = 130 GeV}, \href{https://doi.org/10.1103/PhysRevLett.86.402}{Phys. Rev. Lett. 86 (2001) 402}, [\href{https://arxiv.org/abs/nucl-ex/0009011}{\tt{nucl-ex/0009011}}].

\bibitem{SSA}
S. S. Adler et al. (PHENIX Collaboration), \emph{Elliptic Flow of Identified Hadrons in $Au + Au$ Collisions at $\sqrt{s_{\text{NN}}}$ = 200 GeV}, \href{https://doi.org/10.1103/PhysRevLett.91.182301}{Phys. Rev. Lett. 91 (2003) 182301}, [\href{https://arxiv.org/abs/nucl-ex/0305013}{\tt{nucl-ex/0305013}}].

\bibitem{KA}
K. Aamodt et al. (ALICE Collaboration), \emph{Higher Harmonic Anisotropic Flow Measurements of Charged Particles in Pb-Pb Collisions at $\sqrt{s_{\text{NN}}}$ = 2.76 TeV}, \href{https://doi.org/10.1103/PhysRevLett.107.032301}{Phys. Rev. Lett. 107 (2011) 032301}, [\href{https://arxiv.org/abs/1105.3865}{\tt{1105.3865}}].

\bibitem{GA}
G. Aad et al. (ATLAS Collaboration), \emph{Measurement of event-plane correlations in  $\sqrt{s_{\text{NN}}}$ = 2.76 TeV lead-lead collisions with the ATLAS detector}, \href{https://doi.org/10.1103/PhysRevC.90.024905}{Phys. Rev. C 90 (2014) 024905}, [\href{https://arxiv.org/abs/1403.0489}{\tt{1403.0489}}].

\bibitem{JA}
J. Adam et al. (ALICE Collaboration), \emph{Correlated Event-by-Event Fluctuations of Flow Harmonics in Pb-Pb Collisions at $\sqrt{s_{\text{NN}}}$ = 2.76 TeV}, \href{https://doi.org/10.1103/PhysRevLett.117.182301}{Phys. Rev. Lett. 117 (2016) 182301}, [\href{https://arxiv.org/abs/1604.07663}{\tt{1604.07663}}].

\bibitem{HM}
Harvey B. Meyer, \emph{Calculation of the shear viscosity in SU(3) gluodynamics}, \href{https://doi.org/10.1103/PhysRevD.76.101701}{Phys. Rev. D 76 (2007) 101701(R)}, [\href{https://arxiv.org/abs/0704.1801}{\tt{0704.1801}}].

\bibitem{FW}
F. Karsch and H. W. Wyld, \emph{Thermal Green’s functions and transport coefficients on the lattice}, \href{https://doi.org/10.1103/PhysRevD.35.2518}{Phys. Rev. D 35 (1987) 2518}.

\bibitem{GD}
G. Policastro, D. T. Son, and A. O. Starinets, \emph{Shear Viscosity of Strongly Coupled $N = 4$ Supersymmetric Yang-Mills Plasma}, \href{https://doi.org/10.1103/PhysRevLett.87.081601}{Phys. Rev. Lett. 87 (2001) 081601}, [\href{https://arxiv.org/abs/hep-th/0104066}{\tt{hep-th/0104066}}].

\bibitem{GM}
Guy D. Moore, \emph{Shear viscosity in QCD and why it's hard to calculate}, \href{https://arxiv.org/abs/2010.15704}{\tt{2010.15704}}.

\bibitem{GJ}
Jacopo Ghiglieri \textit{et al.}, \emph{QCD shear viscosity at (almost) NLO}, \href{https://doi.org/10.1007/JHEP03(2018)179}{JHEP \textbf{03} (2018) 179}, [\href{https://arxiv.org/abs/1802.09535}{\tt{1802.09535}}].

\bibitem{GP}
L. Giusti and M. Pepe, \emph{Energy-momentum tensor on the lattice: Nonperturbative renormalization in Yang-Mills theory}, \href{https://doi.org/10.1103/PhysRevD.91.114504}{Phys. Rev. D 91 (2015) 114504}, [\href{https://arxiv.org/abs/1503.07042}{\tt{1503.07042}}].

\bibitem{GP2}
M.~Dalla Brida, L.~Giusti and M.~Pepe,
\emph{Non-perturbative definition of the QCD energy-momentum tensor on the lattice,}
\href{https://doi:10.1007/JHEP04(2020)043}{JHEP \textbf{04} (2020), 043}
[\href{https://arxiv.org/abs/2002.06897}{2002.06897}].

\bibitem{GP3}
M.~Bresciani, M.~Dalla Brida, L.~Giusti and M.~Pepe,
\emph{Thermal QCD for non-perturbative renormalization of composite operators,}
\href{https://doi:10.22323/1.430.0364}{PoS \textbf{LATTICE2022} (2023), 364}
[\href{https://arxiv.org/abs/2211.13641}{2211.13641}].

\bibitem{GP4}
M.~Bresciani, M.~Dalla Brida, L.~Giusti and M.~Pepe,
\emph{Progresses on high-temperature QCD: Equation of State and energy-momentum tensor,}
\href{https://doi:10.22323/1.453.0192}{PoS \textbf{LATTICE2023} (2024), 192}
[\href{https://arxiv.org/abs/2312.11009}{2312.11009}].

\bibitem{RW}
A.  Roberge  and  N.  Weiss, \emph{Gauge  Theories  With  Imaginary  Chemical  Potential  and  the Phases of QCD}, \href{https://doi.org/10.1016/0550-3213(86)90582-1}{Nucl. Phys. B \textbf{275} (1986) 734}.

\bibitem{HH}
H. Makino and H. Suzuki, \emph{Lattice energy–momentum tensor from the Yang–Mills gradient flow—inclusion of fermion
fields}, \href{https://doi.org/10.1093/ptep/ptu070}{PTEP (2014) 063B02}, [\href{https://arxiv.org/abs/1403.4772}{\tt{1403.4772}}], [Erratum: \href{https://doi.org/10.1093/ptep/ptv095}{PTEP (2015) 079202}].

\bibitem{AB}
A. Bazavov \textit{et al.}, \emph{Equation of state in (2+1)-flavor QCD}, \href{https://doi.org/10.1103/PhysRevD.90.094503}{Phys. Rev. D 90 (2014) 094503}, [\href{https://arxiv.org/abs/1407.6387}{\tt{1407.6387}}].




\end{thebibliography}
\end{document}